\documentclass[12pt]{article} 

\usepackage[utf8]{inputenc} 

\usepackage{geometry} 
\usepackage{setspace}

\usepackage{graphicx,url} 
\usepackage{caption}
\usepackage{sidecap}
\usepackage{aas_macros}

\usepackage{booktabs} 
\usepackage{array} 
\usepackage{verbatim} 
\usepackage{amssymb}
\usepackage{amsmath}

\usepackage{fancyhdr} 
\title{\vspace{-8ex}\textbf{How much dark matter really matters?}\vspace{-1ex}}
\author{\vspace{-2ex}
\textbf{Dr.~Jenny Wagner} \\[1ex] 
\small{wagner@asiaa.sinica.edu.tw} \\[-1ex]
\small{Academia Sinica Institute of Astronomy and Astrophysics,}\\[-2ex] 
\small{No.1, Sec.~4, Roosevelt Rd, Taipei 106319, Taiwan, R.O.C} \\[-1ex]
\small{Helsinki Institute of Physics, P.O. Box 64, FI-00014 University of Helsinki, Finland} \\[-1ex]
\small{\textbf{\url{https://thegravitygrinch.blogspot.com}}}}

\begin{document}
\maketitle
\thispagestyle{empty}
\begin{singlespace}
\vspace{-5ex}
\begin{center}
\textit{Essay written for the \\ Gravity Research Foundation 2026 Awards for Essays on Gravitation.}
\end{center}
\end{singlespace}
\begin{figure}[h!]
\begin{center}
\includegraphics[width=0.49\textwidth]{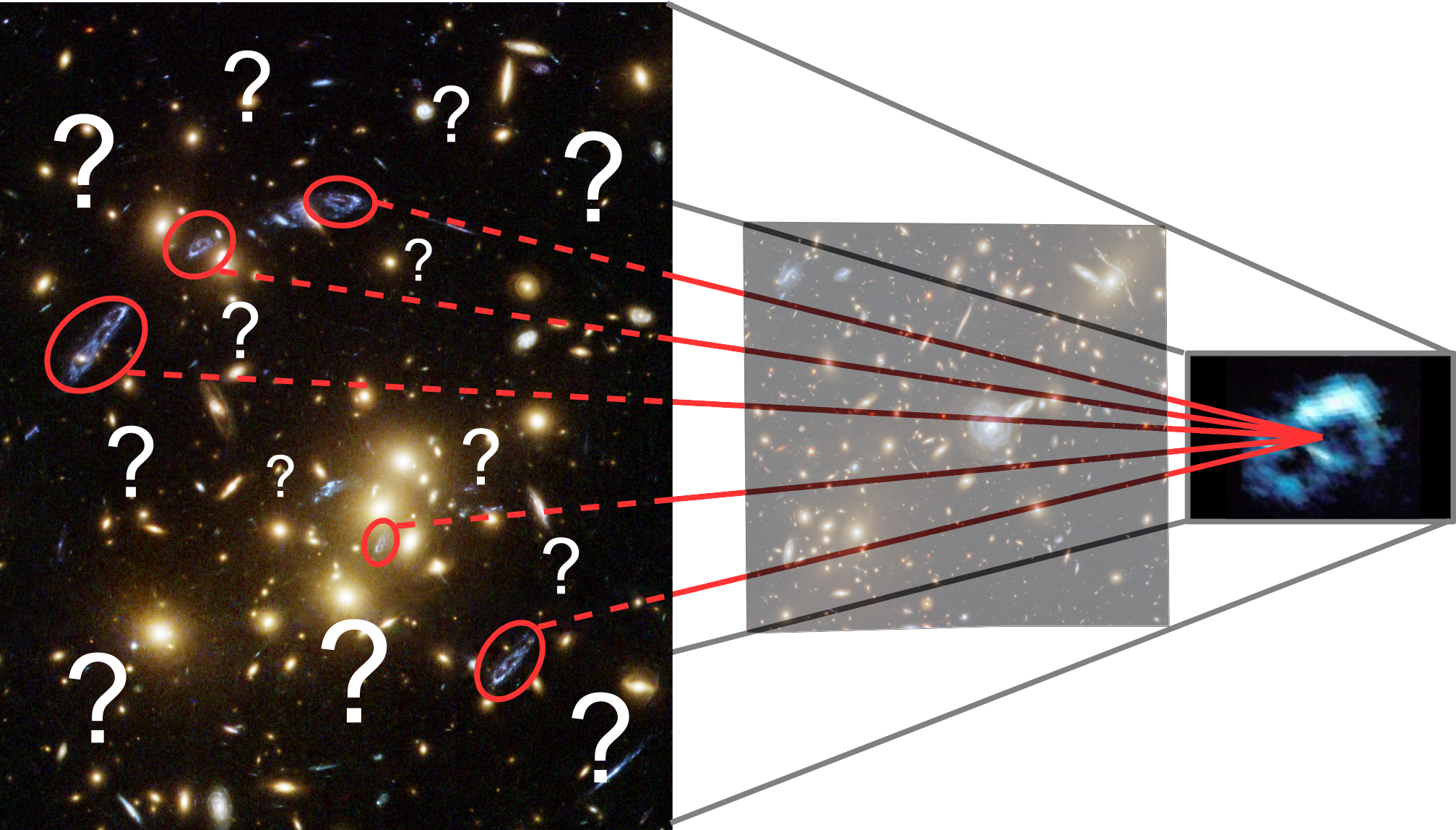}
\end{center}
\end{figure}
\vspace{-4ex}
\abstract{
\noindent 
Strong gravitational lensing is a key probe to trace dark matter. 
It assumes that mass curves spacetime so that light from a background source is deflected on its way to the observer. 
If dark matter contributes the major part to a massive cosmic structure, reconstructing the latter from strong-lensing observables allows us to infer characteristics of dark matter. 
Standard reconstructions fit a pre-defined mass-density model to the data. 
In this essay, I show how these mass models over-estimate the dark-matter contents of light-deflecting masses. Eliminating these models from the reconstruction reveals that observations directly constrain \emph{local} properties of light-deflecting masses. 
How much dark matter is really needed in strong-gravitational-lensing effects and how much do we make up by our model choices? 
} 

\clearpage
\pagenumbering{arabic} 
\subsection*{Modelling light deflections by massive cosmic structures}
Dark matter was introduced in our concordance cosmology to solve the missing-mass problems in the cosmic microwave background and in gravitationally bound structures, \cite{bib:DeSwart2017}. 
To do so, it has to provide 85\% of the total cosmic matter content. 
Yet, it hardly interacts electromagnetically so that we still lack a direct detection to observe its distribution. 
Fortunately, it participates in gravitational interactions and large dark-matter densities curve spacetime as described by General Relativity, \cite{bib:Cirelli2024}.
Strong gravitational lensing is thus an important tool to trace dark matter via the light deflection it causes as part of cosmic structures. 
It only relies on General Relativity to reconstruct the light-deflecting mass distributions, also called ``gravitational lenses'', without the need for hydrostatic or dynamic equilibria like other methods. 

\begin{figure}[ht!]
\centering
\includegraphics[width=0.49\textwidth]{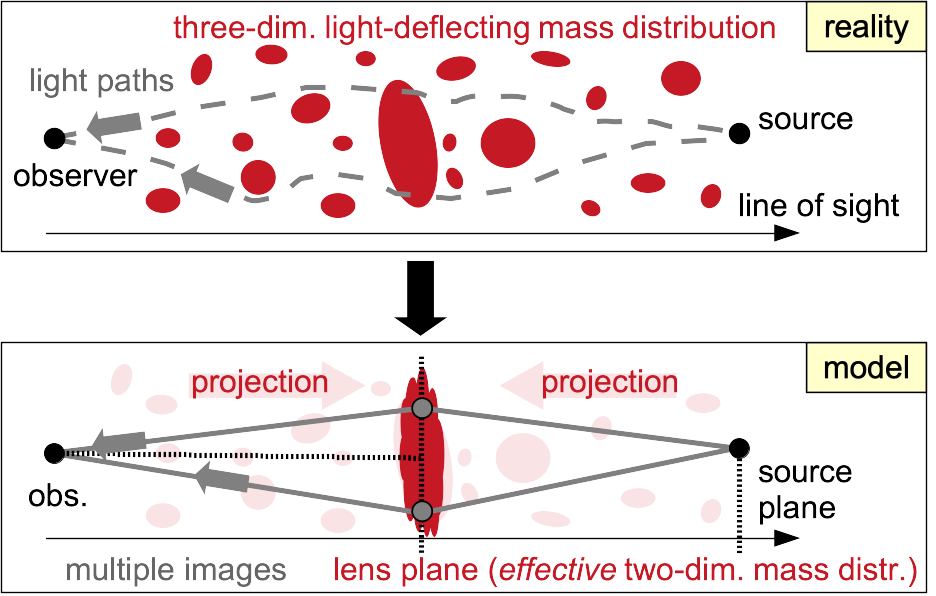}\hspace{4ex}
\includegraphics[width=0.42\textwidth]{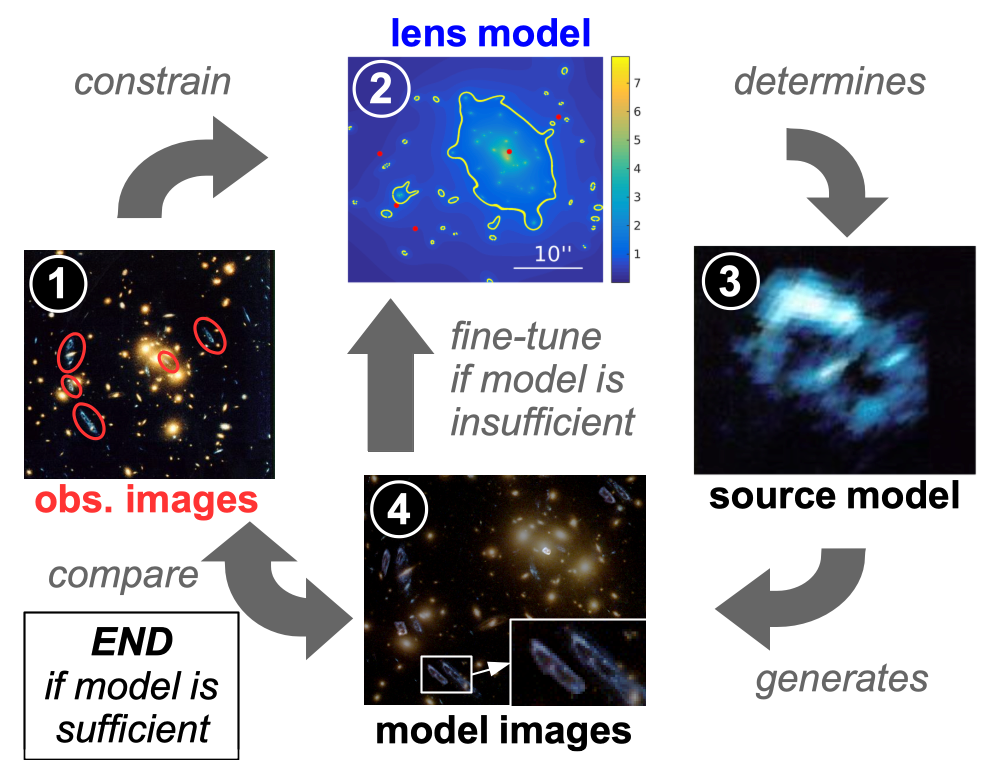}
\caption{Left: Modelling strong gravitational lensing. Right: Standard way to reconstruct an effective mass density map of all light-deflecting masses between us and the background source projected into the plane where the main gravitational lens is located.}
\label{fig:lensing_intro}
\end{figure}

But observables to reconstruct a light-deflecting mass distribution are usually sparse -- typically about 100 multiple images per galaxy cluster and three per galaxy. 
Moreover, we usually assume that additional light-deflecting masses between us and the light-emitting sources are mostly concentrated around the galaxy cluster or the galaxy producing the multiple images. 
We can thus reconstruct a two-dimensional mass-density map at the distance of the main gravitational lens and assume that additional light deflectors cause only small perturbations.
Closer to reality, we better interpret the two-dimensional mass density as an \emph{effective} reconstruction with all light-deflectors projected into a single lens plane.
Fig.~\ref{fig:lensing_intro} (left) from top to bottom depicts this modelling scenario.
Fig.~\ref{fig:lensing_intro} (right) then shows the standard approach to reconstruct this two-dimensional mass map based on the assumption of a global mass density profile whose parameters are inferred from a fit to all observables of the multiple images that we found.

While a fitted mass-density profile yields one possible self-consistent reconstruction of the gravitational lens and the source(s), the solution is by far not unique, as first found by \cite{bib:Falco1985} and \cite{bib:Gorenstein1988}. 
Parameter fits obeying the principle in Fig.~\ref{fig:lensing_intro} (right) take days to weeks to retrieve a large set of possible reconstructions with quite different mass density maps.
Only those incompatible with observed luminous matter distributions can be excluded. 
Hence, if the model-fitting reconstructions are so degenerate, what do we really know about the light-deflecting masses, particularly the dark-matter contribution? 


\subsection*{Separating data-based evidence from model assumptions}

To answer this question, we need to eliminate the lens model from the reconstruction process and determine the information about the light-deflectors that is directly contained in the observables of the multiple images. 
This is also the maximum information that all model fits will agree upon. 

The strong-lensing formalism is constructed by Fermat's principle and correlates \emph{local} differences in the \emph{overall} light-deflection between the light paths, \cite{bib:Tessore, bib:Wagner4, bib:Wagner6, bib:Wagner7}:
It relates arrival-time differences of light pulses deflected into different multiple images to
differences in their two-dimensional light-deflecting gravitational potential and in their geometric paths (Fig.~\ref{fig:lensing_intro}, left bottom).
Being able to measure arrival-time differences for time-varying sources like supernovae or quasars, the \emph{average} mass density inside the area enclosed by the multiple images is constrained. 
Analogously, the total mass inside an Einstein ring can be calculated.
To create such a ring-like multiple-image configuration, a circular two-dimensional mass-density distribution needs to align with a light-emitting source behind its centre along our line of sight.
Apart from such special cases, the data only determines properties of the light deflector at the data positions.
Hence, inserting observables into the equations of the formalism alone constrains \emph{local} lens properties, \cite{bib:Wagner_summary}. 
Including different mass-density profiles to fit the data, the disagreement in their reconstructed mass maps stems from their differing dark-matter distributions in regions devoid of data, see e.g.~\cite{bib:Liesenborgs2012}. 
Yet, all reconstructions agree in the local lens properties directly derived from the formalism.

So, without additional assumptions about the relationship between luminous and dark matter, \emph{all} mass-density profiles that agree with the local lens properties derived from the formalism alone are viable explanations of a multiple-image configuration. 
These findings imply that the dark-matter content of a gravitational lens can only be constrained at the positions of the observables.

\subsection*{Simple mass-density profiles with large lensing masses}

Historically, the first mass-density profiles were smooth circular ones that could fit the observables to the precision achievable in the 2000s. 
Yet, the amount of dark matter in these mass maps seemed unrealistically high.
The complexity of the mass-density models was increased introducing elliptical symmetry or even an amorphous profile of arbitrary shape with clumpy substructures. 
In this way, the dark-matter content in lenses was reduced to more consistent values while keeping the lensing efficiency at the level to explain the lensed giant arcs observed, \cite{bib:Meneghetti2007}. 
Strong lensing became a probe for substructures in galaxy clusters or a detector for satellite galaxies. 
Since the advent of $N$-body simulations to generate mock configurations, discrepancies between observations and simulations have yielded a deeper understanding of possible distributions of lensing masses, particularly along the line of sight, \cite{bib:Despali2018, bib:Meneghetti2017}. 
They also enabled us to investigate new dark-matter properties like self-interactions or varying velocities. 
Comparing summary statistics of observables to simulated ones has already constrained these properties \cite{bib:Vegetti2024}.
Recently, it also raised new questions whether observations of strong-lensing effects caused by galaxies in clusters are too abundant to be compatible with simulated clusters within the $\Lambda$-cold-dark-matter concordance cosmology, \cite{bib:Meneghetti2023}. 
But simulations only probe a limited range of possible structure evolutions and the strong-lensing formalism including lens models suffers from a great deal of degeneracies, too.
So, the inferred dark-matter distributions and properties remain vague or very dependent on prior assumptions/ simulation calibrations. 
Skepticism has increased whether these dark-matter clumps exist that are predicted by lens-model-based mass-density maps but lack any luminous counterparts or even an astrophysical motivation. 
Being there just to reproduce a certain configuration of multiple images is not enough, \cite{bib:Limousin2024}.

\subsection*{Local lens properties as data-based information}

\begin{figure}[h!]
\centering
\includegraphics[width=0.49\textwidth]{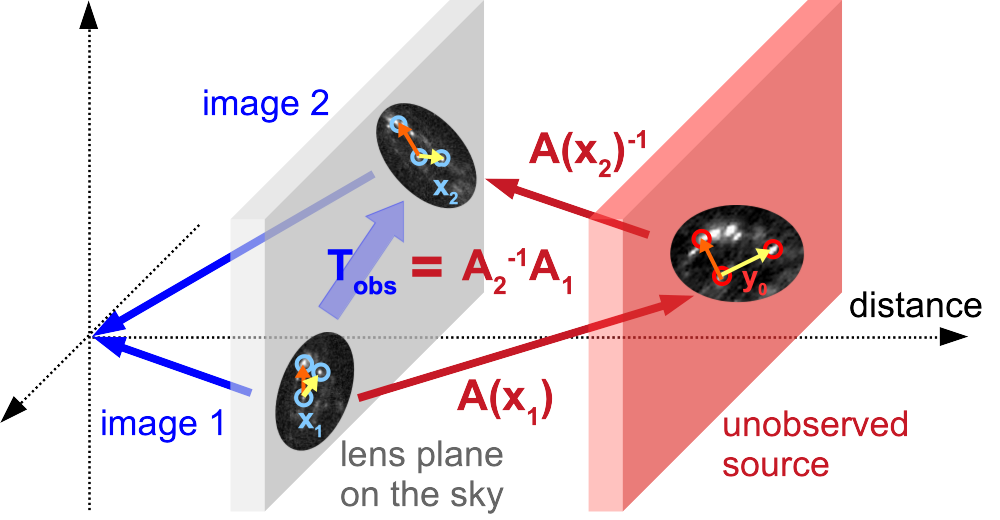}\hspace{2ex}
\includegraphics[width=0.47\textwidth]{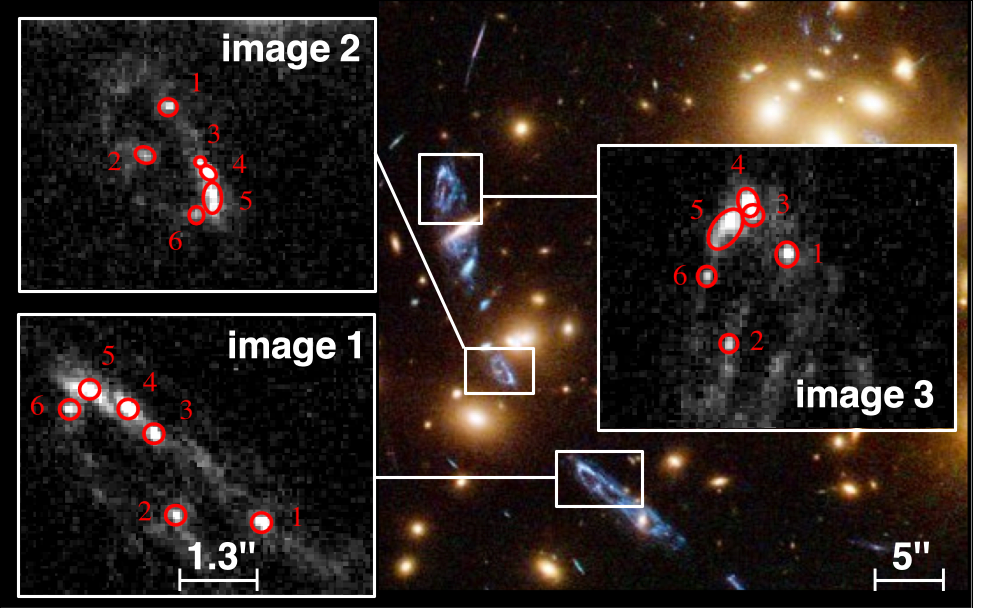}
\caption{Left: Method to infer local lens properties from the surface brightness of the multiple images by setting up a linear transformation T containing local lens properties of Eq.~\eqref{eq:A} (marked in red) and equating it to T set up from observables in the multiple images (marked in blue). Right: Example to set up T by matching corresponding features for three multiple images (marked by red circles, numbers 1-6 indicate the matching).}
\label{fig:method}
\end{figure}

Reducing the evaluation of strong-lensing effects to the local lens properties, our formalism summarised in \cite{bib:Wagner_summary} starts with the fact that the two-dimensional lensing mass density is approximately constant in the vicinity of a multiple image position $x$ on the sky.
How far this vicinity reaches depends on the individual lens.
It can be limited to an area even smaller than a multiple image or extended beyond the images, as detailed below. 
The local lens properties at an image position $x$ are an overall scaling due to the local two-dimensional mass density, $1-\kappa(x)$, and a shearing matrix that captures the local distortion of the surface brightness profile of the source galaxy, the ``reduced shear'' due to masses close to the light path. In combination, they form the distortion matrix
\begin{equation}
\mathrm{A}(x)=(1-\kappa(x)) \left( \begin{matrix} 1 - g_1(x) & - g_2(x) \\ - g_2(x) & 1 + g_1(x) \end{matrix} \right) \;.
\label{eq:A}
\end{equation}
As Fig.~\ref{fig:method} (left) shows, projecting an image~1 to the source plane via its distortion matrix $\mathrm{A}(x_1)$ and then, back-projecting the source onto an image~2 via $\mathrm{A}^{-1}(x_2)$, we can set up a linear transformation $\mathrm{T}$ of local lens properties that maps image~1 onto image~2. 
We can also set up T purely based on observables by matching corresponding features in the surface brightness profiles of the two images with each other, see the numbered red circles Fig.~\ref{fig:method} (right) for an example.
Equating T based on observables, T$_\mathrm{obs}$, with T$=\mathrm{A}^{-1}(x_2)\mathrm{A}(x_1)$ of local lens properties, we can \emph{uniquely} extract $f_{12} \equiv (1-\kappa(x_1))/(1-\kappa(x_2))$, $g_1(x_1)$, $g_2(x_1)$, $g_1(x_2)$ and $g_2(x_2)$.

While these quantities are free of the degeneracies occurring in the lens-model-based approaches, their information is sparse.
Their uncertainties are of the order of 10\%, \cite{bib:Wagner2}, and larger than those of model-based ones. 
But, the models rely on assumptions that need not be true and can lead to biases, see e.g.~\cite{bib:Schneider2013} for such a case.
Different models may even contradict each other. 
The local lens properties are the same in all models and even determined in the same way for all sets of multiple images on galaxy- or cluster-scale. 
They also show how observed brightness features directly translate into local lens properties.

\subsection*{Local lens properties constrain dark-matter properties}

Determining the extent of the region of approximately constant mass density with our approach allows us to infer the local smoothness scale of dark matter. 
As our approach only analyses the areas of the multiple images and around them, this criterion directly probes the smoothness of all matter along these lines of sight. 
It is thus a more direct probe than globally modelling the total mass distribution and subtracting the luminous part. 
Moreover, this approach is independent of any prior assumptions about a specific dark-matter model,  like in \cite{bib:Natarajan2026} 
\noindent
\begin{minipage}[l]{0.37\textwidth}
\vspace{2ex}
\includegraphics[width=\linewidth]{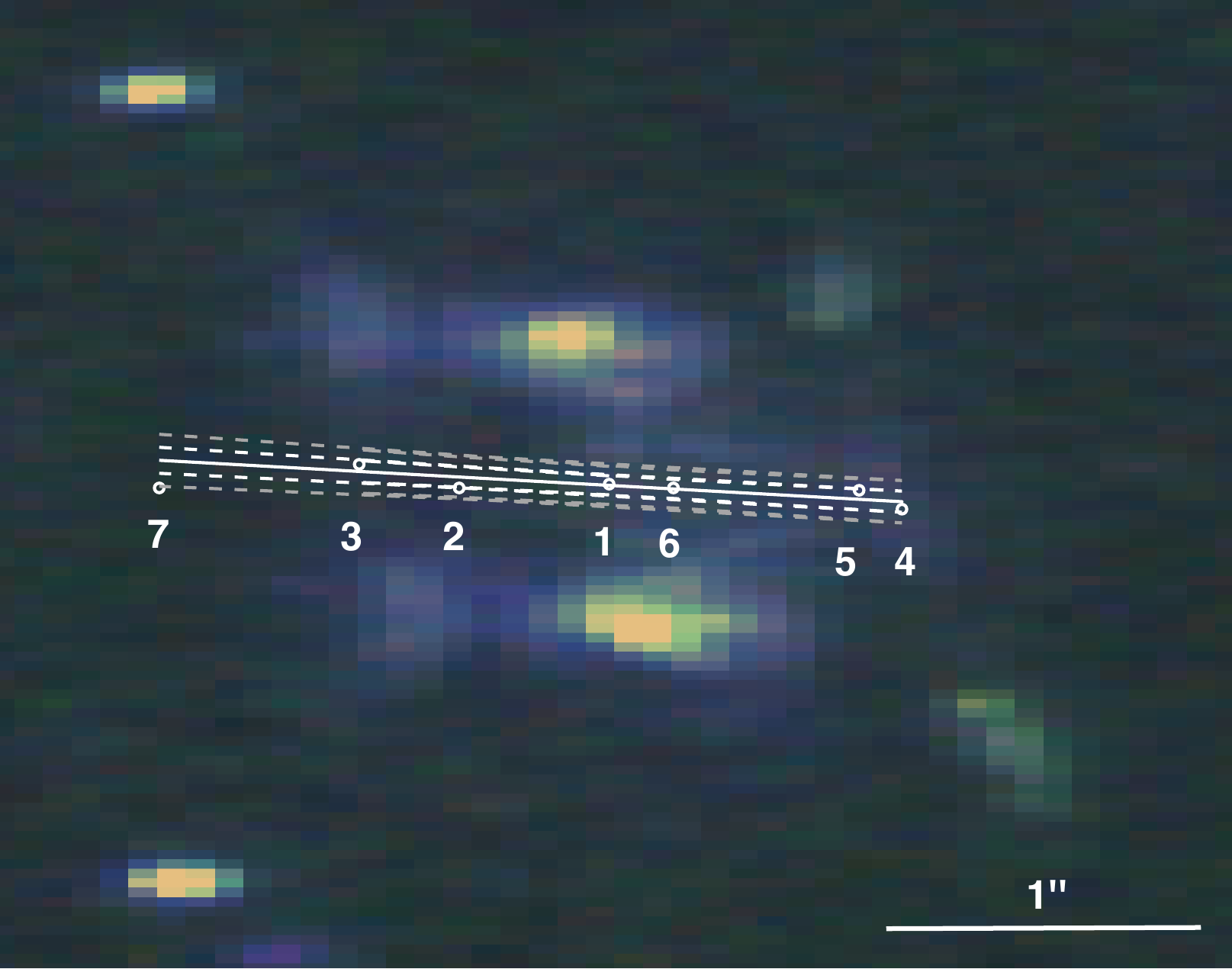} 
\captionof{figure}{Smoothness scale of ca.~1'' (6~kpc) constrained by two mirroring images in the galaxy cluster RM J223013.}
\label{fig:eval_props}
\end{minipage}
\hspace{2ex}
\begin{minipage}[r]{0.47\textwidth}
\vspace{2ex}
and a macro lens model, like in \cite{bib:Dai2020}. 
For the recently discovered galaxy cluster RM~J223013, we measured an extent of the local smoothness scale between two multiple images of at least 6~kpc (see Fig.~\ref{fig:eval_props}), \cite{bib:Griffiths2021}.
In the galaxy cluster Abell~3827, this region was smaller than the area covered by a multiple image, \cite{bib:Lin2023}. 
So our approach could only be applied to parts of the multiple images.
Yet, these images are larger than those 
\end{minipage}

\vspace{2.5ex}
\noindent
in RM J223013 and their distance is smaller to us, so that the smoothness scale in Abell~3827 is also around 5~kpc. 
Both smoothness scales are lower bounds, as there is no sign of dark-matter clumps disrupting the continuity of the lensing. 
\begin{figure}[ht]
\centering
\includegraphics[width=0.88\textwidth]{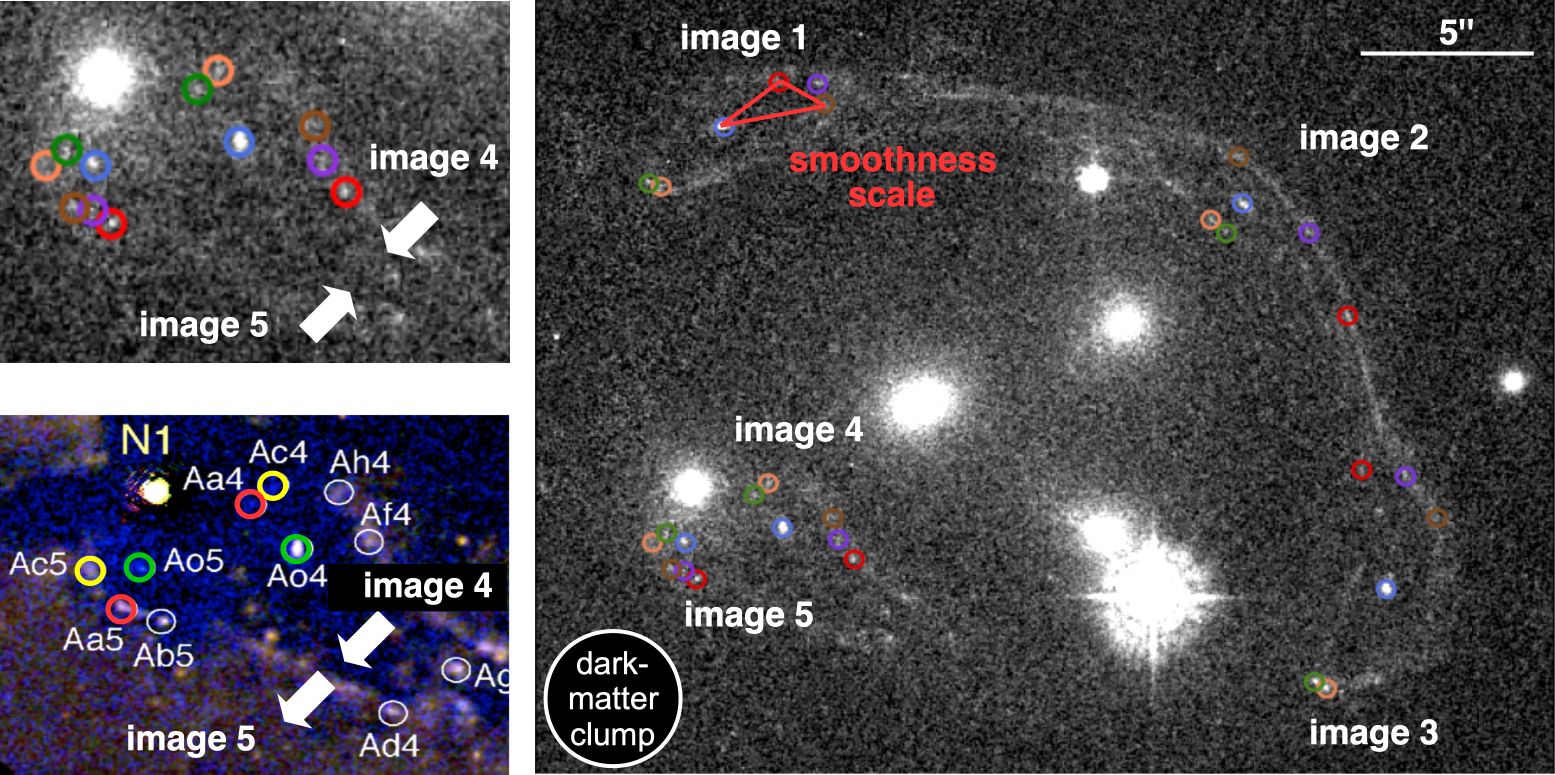}
\caption{Left: Ambiguous feature matching (coloured circles indicate matched brightness features): mirror images (top) or twin-images (bottom) requiring different external shears in Abell~3827. Right: Hubble-Space-Telescope observation of Abell~3827 with five detailed images, the position of the dark-matter clump (black circle) needed when modelling the shear for the twin-images and the smoothness scale between features in image~1 (red triangle) that also exists for the same features in the other images. There are two more point-like images from the same source in the centre (not shown).}
\label{fig:Abell3827}
\end{figure}

A first candidate for a dark-matter clump without a stellar counterpart was found close to galaxy M94, \cite{bib:Benitez2024}.
While a follow-up analysis confirmed that the clump contains hardly any stars, the study found neutral hydrogen gas with a mass two orders of magnitude larger than the stellar one, \cite{bib:Anand2025}. 
So it still seems unlikely that dark clumps without any luminous counterparts found in lens-model-based mass-density maps are real. 
Such objects are often put at the borders of lens-model reconstructions, like in Abell~3827 (see Fig.~\ref{fig:Abell3827}, right). 

Modelling the global mass-density map of Abell~3827, a dark clump appears close to a pair of images that requires a strong shear, if the pair is interpreted as two ``twin images", as shown in Fig.~\ref{fig:Abell3827} (left, bottom). 
Adding constraints on the global mass map from another multiple-image system close to the dark clump shifts the clump away.
To still generate the required shear for the pair of twin-images, a stretched-out version of the dark clump reappears at the other end of the reconstruction area along the same relative direction with respect to the twin-images, \cite{bib:Mohammed2014}. 
But the matching of the brightness features used to map the two images onto each other is ambiguous.
It is also possible to match the features such that the pair consists of two images mirroring each other, as shown in Fig.~\ref{fig:Abell3827} (left, top). 
Then, the dark clump does not appear in the lens-model-based reconstruction because no strong shear is required to generate this rather common image configuration.

Extending the mass-density reconstruction to include weakly lensing masses at larger distances can reveal if the strong shear of the first matching can have an astrophysical origin in the lens plane. 
Another option to explain the dark clump(s) is that Abell~3827 with its line-of-sight extent of about 10~Mpc is a thick lens compared to its distance of about 400~Mpc to us. 
For other lenses, their extent is much smaller compared to their distance from us. 
For instance, RM~J223013 is ca.~1,300~Mpc away from us, but expected to have the same line-of-sight extent as Abell~3827. 
Hence, non-linear lens-lens couplings between the individual galaxies in Abell~3827 could cause the strong shear of the first matching. 
Opening up for a full three-dimensional lens reconstruction instead of the effective two-dimensional approach allows for relative rotations between multiple images beyond the local lens properties in Eq.~\eqref{eq:A}. 
This scenario can also explain the slight relative rotations between images~1 to 3 of the same source also observed in Abell~3827. 

A three-dimensional reconstruction will also clarify if the cluster-size dark-matter halo currently used in the lens-model-based reconstructions is really necessary, as claimed in \cite{bib:Chen2020} that performed a two-dimensional effective reconstruction. 
The latter work investigated in how far the multiple-image configuration consisting of 7 images from the same background source can be explained by the luminous matter of the cluster alone, i.e.~the stellar, intra-cluster gas, and X-ray-emitting hot gas mass densities. 
The authors did not succeed because the required mass-to-light ratio for the intra-cluster gas was too high.
Moreover, the ellipticity of the intra-cluster gas mass density and of one of the four central galaxies were also too large compared to the ellipticity of the light distribution observed. 
So it was concluded that a decoupled, dominating dark-matter halo could generate the multiple-image configuration and, at the same time, allow the properties of the intra-cluster gas and the central galaxies to be fixed by their luminous observables. 
While this conclusion might be \emph{sufficient}, it is by far not \emph{necessary}, as supported by our findings and the arguments given above.
Therefore, a fresh look into a three-dimensional mass reconstruction of Abell~3827 will reveal where additional mass to the already observed one is actually required and can potentially be provided by luminous structures not yet taken into account in the reconstruction of the strong-lensing effects. 

\begin{SCfigure}
\includegraphics[width=0.44\textwidth]{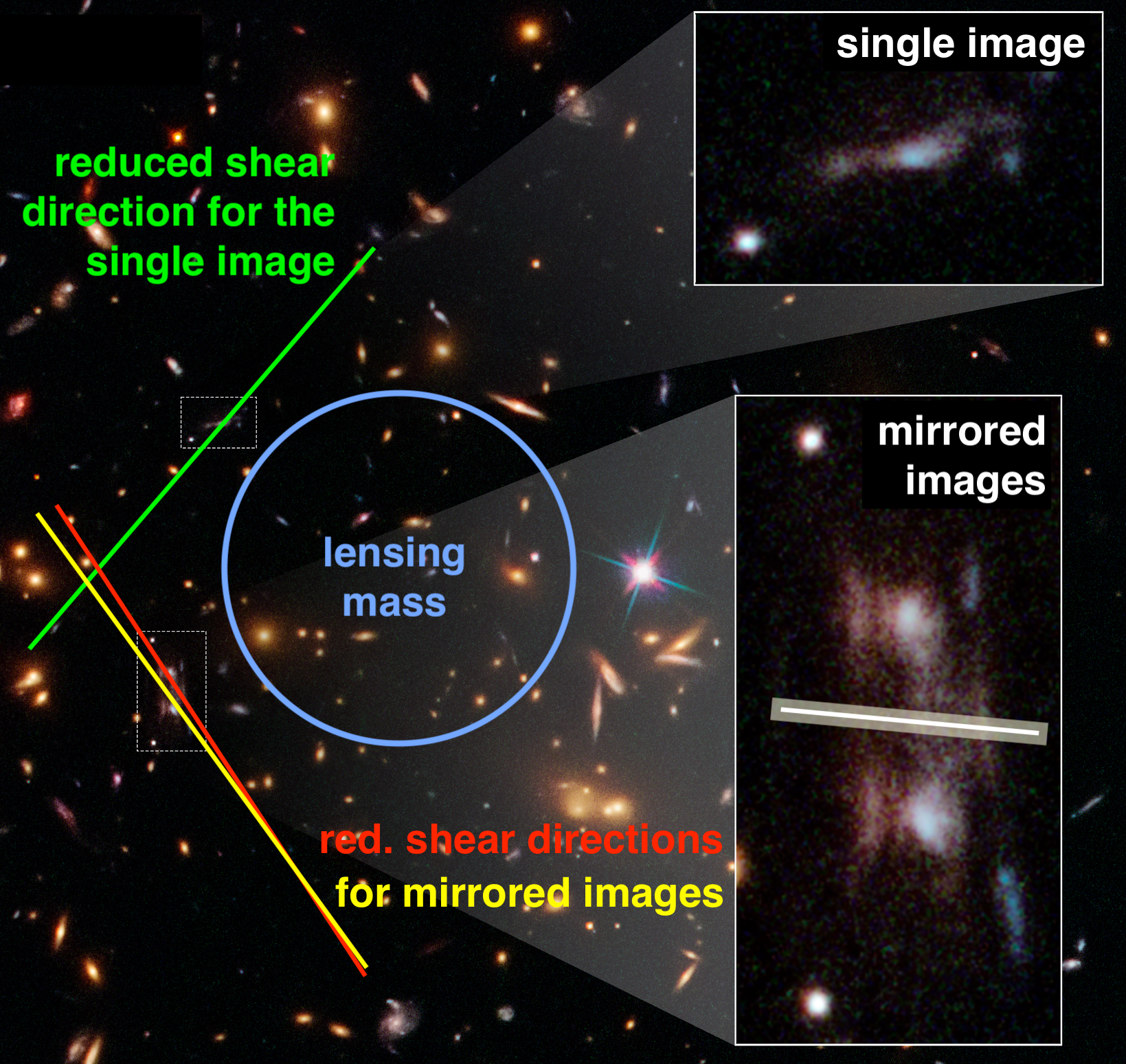}
\vspace{-2ex}
\caption{Left: Using the local reduced shear directions (coloured lines) from the three multiple images lensed by RM~J223013 to find the mass causing this so-called ``cusp'' image configuration (blue circle). Right: Details of the individual images show that each contains seven brightness features used to determine the reduced shear at each image position, and the smoothness scale of dark matter for the mirrored images, as shown in Fig.~\ref{fig:eval_props}.}
\label{fig:Hamilton}
\end{SCfigure}

That our local lensing approach can also help to locate lensing masses in a global mass-density map was demonstrated in \cite{bib:Griffiths2021}:
Since only three multiple images of a single background galaxy have been discovered in RM J223013 so far, any lens-model fit does not shed much light into the mass map of this cluster. 
However, using the locally inferred reduced shear directions at all three images, we estimated the location of the lensing mass from the relative shear orientations, as shown in Fig.~\ref{fig:Hamilton}. 
As the position of this lensing mass does not coincide with the centre of the cluster as estimated by the red-sequence of galaxies, we presumed the cluster to be a merger, which was recently confirmed by an independent study, \cite{bib:Ebeling2025}. 

Even on the scale of constraining substructures in galaxy clusters inside the light paths of the multiple images, lens-model-based mass-density reconstructions suffer from a degeneracy between the surface brightness profile of the source and small-scale dark matter clumps, \cite{bib:Wagner7}. 
Since we lack knowledge of the source brightness profile and the smoothness of dark matter on small scales, we can either assume a rather smooth surface brightness profile for the source and attribute fluctuations to the lensing or, vice versa assume a smooth lensing effect on a more perturbed galaxy, see, e.g.~\cite{bib:Eid2026} for a recent model-based source reconstruction approach. 
Depending on the assumed source and lens models, artefacts can occur, either due to a low signal-to-noise (as is the case for Abell~3827)  or due to the regularisation assumption of the individual models (see \cite{bib:Ephremidze2025}).
Our local lensing approach also contributes to identify such artefacts from the start by construction, as the relative distances between the brightness features in the multiple images naturally set the minimum length scale on which observables can be used to constrain source or lens properties.
They set a data-based pixelisation of the source and thus fix the maximum resolution of the local lens reconstruction as well.
So it is possible to avoid the introduction of dark-matter artefacts in the lens, as well as one can identify artefacts in the source reconstruction for noisy surface brightness profiles of the multiple images. 
The latter was also demonstrated for a galaxy-scale lens in the radio band, \cite{bib:Wagner_quasar}, when our approach yielded contradicting local lens properties, so that all possible brightness-feature matchings between the images were deemed to be wrong. 
Moreover, \cite{bib:Wagner3} showed in how far observables in giant arcs are able to constrain properties of smaller-scale satellites to lensing galaxies. 
The results yield a deeper understanding of the degeneracies that lead to the detection of non-existing perturbers and the broad range of possible properties of existing perturbers in a more efficient way than using a lens-model-based method with simulations.

\subsection*{Local lens properties test fundamental principles}

Since our approach does only rely on the strong-lensing formalism without making specific assumptions about the light-deflecting structures, we can also test fundamental principles with observations. 
One example is the wavelength independence of gravitational lensing, another one is the scale-freeness of Einsteinian and Newtonian gravity that should lead to scale-free local lens properties, unless some modified gravitational law holds.
Observations from multiple filter bands are available, for instance from the Hubble or James Webb Space Telescopes, so we investigated two different strong-lensing examples on galaxy-cluster scale for potential wavelength dependencies, \cite{bib:Lin2022}. 
Within the error bounds of about 10\% the local lens properties across all filter bands coincided, confirming the wavelength independence. 

For the test of scale-freeness, we need to compare the local lens properties from lenses of different size and mass scales. 
Since we have only analysed a single galaxy-scale lens so far, which even showed an artefact due to a low signal-to-noise ratio, this test is still to be performed more thoroughly.
The results obtained so far, however, look promising that local lens properties also agree across different scales.
They are already in good agreement, i.e. of the same order of magnitude, for three different observed and one simulated galaxy-cluster lenses.

\subsection*{Conclusion}

As this essay showed, we can directly connect observables in strong-lensing configurations with the light-deflecting \emph{local} lens properties that are \emph{necessary} to create these lensing effects.
This yields the maximum information contained in the data.
As detailed in \cite{bib:Wagner3,bib:Wagner4, bib:Wagner6, bib:Wagner7} a deeper and clear understanding of all degeneracies occurring in the strong-lensing formalism can be gained from the derivation of this information. 
So, after 30 years of studying strong-lensing effects, we finally learn that mass-density profiles full of dark matter are so successful in reproducing the lensing observables because there is an ample volume devoid of data that can be filled with almost arbitrarily distributed dark-matter mass densities. 
Even at the positions of the multiple images fundamentally unbreakable degeneracies exist unless we find a way to trace dark matter (or observe the source without a lensing effect).
One may argue that these degeneracies could be broken with an ensemble of strong-lensing configurations or with complementary data. 
While both approaches alleviate the degeneracies, they do not fully break them.
The former contains potential selection biases when assembling a statistically representative set of examples and, for the latter, other cosmological probes suffer from the undetectability of dark matter in very similar ways. 
Hence, how much dark matter we really need is a question that can be answered from joining data-based ``patch-work'' information as outlined here and, where necessary, add supplementary models but systematically test their impact on the interpretation of the observables. 
Most importantly, having reached the data-driven era, astrophysics advances more efficiently by leaving knowledge gaps to be filled with future data than holding on to a bag of model assumptions whose predictions have been difficult to clearly falsify or support for decades, \cite{bib:Aluri2023}.
A promising path forward goes back to the foundations of General Relativity, as estimated in \cite{bib:Buchert2009}. 
If the sun curves space to double the deflection angle of background stars compared to the Newtonian formulation, the local curvature of galaxies and clusters may not be incorporated correctly in the quasi-Newtonian strong-lensing formalism. 
Maybe, the missing mass eluding a direct detection so far is actually a missing local curvature perturbation. 
With the deeper understanding of the degeneracies pointed out here, this interpretation becomes plausible now. 

\newpage
\noindent
\textbf{Acknowledgements} \\
I gratefully acknowledge the support of an O'Shaughnessy Ventures Fellowship.

\subsection*{References}
\begingroup
\renewcommand{\section}[2]{}%

\bibliographystyle{spmpsci}      
\bibliography{darkmatter}   
\endgroup

\end{document}